\documentclass[twocolumn,showpacs,preprintnumbers]{revtex4}

\usepackage{graphicx}
\usepackage{amsmath}
\usepackage{array}
\usepackage{tabularx}

\usepackage{natbib}

\def\be{\begin{equation}}
\def\ee{\end{equation}}
\def\ba{\begin{eqnarray}}
\def\ea{\end{eqnarry}}
\def\bal#1\eal{\begin{align}#1\end{align}}

\newcolumntype{W}{>{\raggedright\arraybackslash}X}
\newcolumntype{Y}{>{\raggedleft\arraybackslash}X}
\newcolumntype{Z}{>{\centering\arraybackslash}X}

\begin{document}
\title{Crystallization of classical multi-component plasmas}
\author{Zach Medin}\email{zmedin@physics.mcgill.ca}
\author{Andrew Cumming}\email{cumming@physics.mcgill.ca}
\affiliation{Department of Physics, McGill University, 3600 rue University,
Montreal, QC H3A 2T8, Canada}

\date{\today}

\begin{abstract}
We develop a method for calculating the equilibrium properties of the
liquid-solid phase transition in a classical, ideal, multi-component
plasma. Our method is a semi-analytic calculation that relies on
extending the accurate fitting formulae available for the one-, two-,
and three-component plasmas to the case of a plasma with an arbitrary
number of components. We compare our results to those of Horowitz,
Berry, \& Brown (Phys.~Rev.~E {\bf 75}, 066101, 2007), who use a
molecular dynamics simulation to study the chemical properties of a
17-species mixture relevant to the ocean-crust boundary of an
accreting neutron star, at the point where half the mixture has
solidified. Given the same initial composition as Horowitz et al., we
are able to reproduce to good accuracy both the liquid and solid
compositions at the half-freezing point; we find abundances for most
species within $10\%$ of the simulation values. Our method allows the
phase diagram of complex mixtures to be explored more thoroughly than
possible with numerical simulations. We briefly discuss the
implications for the nature of the liquid-solid boundary in accreting
neutron stars.
\end{abstract}
\smallskip

\maketitle

\section{Introduction}
\label{sec:intro}

During the crystallization of a plasma containing multiple ion
species, the chemical composition of the solid is in general different
from that of the liquid. This type of chemical separation is important
for both white dwarfs \citep{hansen03} and accreting neutron stars
\citep{horowitz07}. The interior of a white dwarf is a mixture of
carbon, oxygen, and traces of other elements, most abundantly neon. As
the star cools, chemical separation leads to the formation of an
oxygen- and neon-rich core. The energy released through the
gravitational settling of the denser core material heats the star and
can delay cooling by several Gyr \citep{isern91}. A neutron star
accretes mostly hydrogen and helium from its companion, but this
material undergoes a series of nuclear reactions, including rapid
proton capture \citep{schatz01} and then electron capture reactions
\citep{gupta07}, to produce a variety of elements. Through accretion
the mixture is pushed deep into the star and solidifies. Recent
numerical simulations have shown that the mixture undergoes chemical
separation during solidification \citep{horowitz07}, possibly forming
a two-phase solid \citep{horowitz09b}. The composition of the liquid
ocean and the structure and composition of the crust have important
implications for a range of observed phenomena. For example, the
resulting thermal conductivity determines the cooling rate of
transiently accreting neutron stars following extended accretion
outbursts \citep{shternin07,brown09}. The mechanical strength of the
crust limits the size of a possible crust quadrupole and therefore
gravitational wave emission \citep{horowitzkadau09}.

Several groups have studied the liquid-solid phase transition and
chemical separation of two- and three-component plasmas in the
classical, ideal limit (i.e., ignoring quantum mechanical effects on
the ions and treating the electrons as a uniform background;
cf. Ref.~\cite{potekhin00}). Early works (e.g.,
Ref.~\cite{mochkovitch83}) studied phase transitions in carbon-oxygen
plasmas, but the approximations used were too crude for application to
the interiors of white dwarfs. Accurate calculations using the mean
spherical approximation in the density-functional formalism were
performed by \citet{barrat88}, who studied carbon-oxygen plasmas, and
by \citet{segretain93}, who studied arbitrary two-component plasmas
with atomic number $Z$ ratios up to 2 (see also
Ref.~\cite{segretain96}, where carbon-oxygen-neon plasmas are
examined). Using Monte Carlo calculations and $Z$ ratios up to 5,
\citet{ogata93} studied arbitrary two- and three-component plasmas and
\citet{dewitt03} studied arbitrary two-component plasmas with a very
accurate measurement of the liquid free energy (see also
Refs.~\cite{iyetomi89,dewitt96}). All of these groups present phase
diagrams as a function of ion abundance, and some
\citep{ogata93,dewitt03} also present fitting formulae for the liquid
and solid free energies. Using these diagrams and fitting formulae,
one can determine the phase transition properties for a two-component
plasma of any ion type and abundance.

These calculations are particularly useful for the interior of a white
dwarf, where there are only two or three dominant elements. But in the
ocean of an accreting neutron star there are around 10-20 elements
with abundances $> 1\%$ \citep{gupta07}, each one with a potentially
important effect on the behavior of the phase transition and chemical
separation of the mixture. The available analytic or numerical results
for this type of system are extremely limited. We are aware of only
one study of phase transitions in plasmas with more than three
components, that of \citet{horowitz07} (see also
Refs.~\cite{horowitz09a,horowitz09b}). These authors used molecular
dynamics simulations to study a 17-component plasma with a composition
similar to that expected at the ocean-crust interface of an accreting
neutron star. Due to the large amount of computing power necessary to
run each simulation, the phase transition properties have so far only
been calculated for one composition.

We present here a method for rapidly calculating the properties of the
liquid-solid phase transition in a multi-component plasma in the
classical ideal limit, for any initial composition and ion types. Our
method is a semi-analytic calculation that relies on extending the
accurate fitting formulae available for the one-, two-, and
three-component plasmas to the case of a plasma with an arbitrary
number of components. We test our method using the one data point
available for a plasma with more than three components, the
calculation of \citet{horowitz07}, and show that it performs very well
in that specific case.

The paper is organized as follows. In Section~\ref{sec:method} we
describe the semi-analytic calculation as it applies to the
one-component plasma (Section~\ref{sub:ocp}), the two-component plasma
(Section~\ref{sub:tcp}), and the multi-component plasma
(Section~\ref{sub:mcp}). In Section~\ref{sec:results} we present our
results for the 17-component mixture of \citet{horowitz07}. We
conclude in Section~\ref{sec:discuss}. The pressure term in the Gibbs
free energy and its effect on the phase transition, the importance of
the deviation from linear mixing for the liquid free energy, and a
simplified derivation of the deviation from linear mixing for the
solid free energy, are discussed in three appendices.

\section{Method}
\label{sec:method}

\subsection{The one-component plasma}
\label{sub:ocp}

We assume in this paper that the system has reached equilibrium, i.e.,
the state of lowest free energy. The validity of this assumption and
non-equilibrium effects such as diffusion and sedimentation will be
discussed in a later paper. We also assume here that the phase
transition happens at constant volume, in which case the equilibrium
configuration of the system is determined by the state with the lowest
Helmholtz free energy, $F = U-TS$. In reality the transition happens
at constant pressure and at minimized Gibbs free energy. The error
introduced by using the constant volume approximation is discussed in
Appendix~\ref{sec:gibbs}. We find that for the mixture considered in
Section~\ref{sec:results}, the abundance in the liquid state of each
ion species is in error by no more than $2\%$. While the percentage
errors in the abundances in the solid state are typically larger by
factors of $\sim 2$-$5$, the absolute errors for each ion species are
similar in either state. (Since this trend holds true for most of the
approximations we make in this paper, we hereafter quote errors in our
approximations only for the liquid abundances.) Note that in
transitions at constant volume, the free energy of the electrons is
identical in the liquid and in the solid and so has no effect on the
properties of the phase transition.

The Helmholtz free energy of the liquid or solid phase of a
one-component plasma (OCP) can be described as a function of only the
number of ions $N$, the temperature $T$, and the Coulomb coupling
parameter $\Gamma \equiv (Ze)^2/(ak_BT) = Z^{5/3}\Gamma_e$. Here $Ze$
is the ion charge, $a$ is the ion separation, and $k_B$ is the
Boltzmann constant; $\Gamma_e \equiv e^2/(a_ek_BT)$ is the electron
coupling parameter, where $a_e = [3/(4\pi n_e)]^{1/3}$ is the mean
electron spacing and $n_e = ZN/V$ is the electron density.

The ideal gas contribution to the free energy of a one-component
plasma $F_{\rm ideal}$ is given by
\bal
f_{\rm ideal} \equiv {}& \frac{F_{\rm ideal}}{Nk_BT} = \ln\left[\frac{N}{V}\left(\frac{h^2}{2\pi m_ik_BT}\right)^{3/2}\right]-1 \nonumber\\
 {}& \quad = 3\ln\Gamma+\frac{3}{2}\ln(k_BT)_{\rm Ry}-1-\ln\frac{4}{3\sqrt{\pi}} \,,
\label{eq:fgOCP}
\eal
where $m_i=Am_p$ is the mass of the ion and $(k_BT)_{\rm Ry} = k_BT
2\hbar^2/(m_iZ^4e^4)$ is the thermal energy expressed in ionic Rydberg
units. The free energy of the liquid phase of a one-component plasma
$F_l^{\rm OCP}$ is well fit for $\Gamma \in [1,200]$ by
\bal
f_l^{\rm OCP}(\Gamma) \equiv \frac{F_l^{\rm OCP}}{Nk_BT} = {}& -0.899172\Gamma+1.8645\Gamma^{0.32301} \nonumber\\
 & -0.2748\ln(\Gamma)-1.4019 \,.
\label{eq:flOCP}
\eal
The previous formula is from the Monte Carlo calculations of
\citet{dewitt03}, with the modification that the ideal gas
contribution to the free energy [Eq.~(\ref{eq:fgOCP})] has been
removed. Other formulae for $f_l^{\rm OCP}$ can be found in
Refs.~\cite{stringfellow90,ogata87,caillol99,potekhin00} (see also
Ref.~\cite{hansen73,farouki93}); for the range of $\Gamma$ we are
concerned with in this paper ($15 \alt \Gamma \le 200$), the
differences between these formulae, and between the numerical data
these formulae are based on, are less than $0.006$.

The free energy of the solid phase of a one-component plasma $F_s^{\rm
OCP}$ is well fit for $\Gamma \in [160,2000]$ by
\bal
\frac{F_s^{\rm OCP}}{Nk_BT} = {}& -0.895929\Gamma+1.5\ln(\Gamma)-1.1703 \nonumber\\
 & -\frac{10.84}{\Gamma}-\frac{176.4}{\Gamma^2}-\frac{5.980\times10^4}{\Gamma^3} \,.
\label{eq:fsfull}
\eal
The previous formula is from \citet{dubin90}; it was derived using a
combination of analytic methods and a fit to the Monte Carlo
calculations of Ref.~\cite{slattery82}. As in the liquid case, we have
modified Eq.~(\ref{eq:fsfull}) from its original form by removing the
ideal gas contribution. Another formula for $F_s^{\rm OCP}/(Nk_BT)$ of
similar accuracy (with less than $0.004$ difference from
Ref.~\cite{dubin90} or the numerical data for $160 \le \Gamma \le
2000$) can be obtained from the molecular dynamics calculations of
Ref.~\cite{farouki93} (see also
Refs.~\cite{pollock73,stringfellow90}). In this paper we neglect the
$\Gamma^{-2}$ and $\Gamma^{-3}$ terms in Eq.~(\ref{eq:fsfull}) and
use the following approximation for $F_s^{\rm OCP}$:
\bal
f_s(\Gamma)^{\rm OCP} \equiv \frac{F_s^{\rm OCP}}{Nk_BT} \simeq {}& -0.895929\Gamma+1.5\ln(\Gamma) \nonumber\\
 & -1.1703-\frac{10.84}{\Gamma} \,.
\label{eq:fsOCP}
\eal
This expression fits the numerical data for $160 \le \Gamma \alt 300$
with an accuracy several times lower than that of
Eq.~(\ref{eq:fsfull}) [differing by up to $0.02$ for $\Gamma \sim
160$]. We use this expression in place of Eq.~(\ref{eq:fsfull}),
however, because it behaves qualitatively better for small $\Gamma$,
as we discuss below.

The free energy difference based on these fits is given by
\bal
\delta f_{\rm fit}^{\rm OCP}(\Gamma) \equiv {}& \,(f_l-f_s)^{\rm OCP} \nonumber\\
 = {}& -0.003243\Gamma+1.8645\Gamma^{0.32301} \nonumber\\
 & -1.7748\ln(\Gamma)-0.2316+10.84/\Gamma \,.
\label{eq:delfOCP}
\eal
In equilibrium the system will be in the state of lowest free energy:
when $\delta f^{\rm OCP} < 0$, the OCP is in the liquid state, and when
$\delta f^{\rm OCP} > 0$, it is in the solid state. When $\delta f^{\rm OCP}
= 0$ there is a phase transition between the liquid and solid
state. This occurs at
\be
\Gamma_{\rm crit} = 178.6
\label{eq:gammacrit}
\ee
in the above equation. Note that if we had used Eq.~(\ref{eq:fsfull})
instead of Eq.~(\ref{eq:fsOCP}) to calculate $\delta f_{\rm fit}^{\rm
OCP}$, we would obtain $\Gamma_{\rm crit} = 175.3$, which is in
agreement with the most accurate estimates currently available for
this value (e.g., $\Gamma_{\rm crit} = 175.0\pm0.4$ in
Ref.~\cite{potekhin00}); our $\Gamma_{\rm crit}$ differs from the true
transition value by about $2\%$.

Equation~(\ref{eq:delfOCP}) is only accurate for $\Gamma \in
[160,200]$. While there are no Monte Carlo or molecular dynamics data
available for $f_l^{\rm OCP}$ when $\Gamma > 200$, \citet{ichimaru87}
have calculated $f_l^{\rm OCP}$ up to $\Gamma=1000$ using the
``improved hypernetted chain'' (IHNC) method. For $\Gamma \in
[200,1000]$, if $f_l^{\rm OCP}$ is given by Ref.~\cite{ichimaru87} and
$f_s^{\rm OCP}$ is given by Eq.~(\ref{eq:fsOCP}), the approximation
\be
\delta f^{\rm OCP}(\Gamma) = 0.09+0.0043(\Gamma-200)
\label{eq:delfOCPhigh}
\ee
fits the free energy difference to within $0.2$. This error is of
similar magnitude to the error in the IHNC method for $\Gamma > 200$
(as extrapolated from comparisons between IHNC approximations and
Monte Carlo calculations at $\Gamma < 200$; see, e.g.,
\cite{iyetomi83,iyetomi92}), and is several times smaller than the
error that would be obtained by a direct application of
Eq.~(\ref{eq:delfOCP}) to the domain $\Gamma \in [200,1000]$.

There are currently no published results (numerical or otherwise) for
$f_l^{\rm OCP}$ above $\Gamma=1000$ or $f_s^{\rm OCP}$ below
$\Gamma=160$. However, we expect $\delta f^{\rm OCP}$ to increase
monotonically with increasing $\Gamma$, not just in $[160,1000]$ but
for all $\Gamma$. In other words, for the OCP the solid state should
always become more stable with respect to the liquid as $\Gamma$
increases, and less stable as $\Gamma$
decreases. Equation~(\ref{eq:delfOCPhigh}) extended out to arbitrarily
large $\Gamma$ remains consistent with this assumption, but
Eq.~(\ref{eq:delfOCP}) extended down to $\Gamma=0$ does not. This is
because $\delta f_{\rm fit}^{\rm OCP}$ decreases with $\Gamma$ for
$\Gamma \in [0,50]$. An even stronger argument against $\delta f_{\rm
fit}^{\rm OCP}$ representing the true free energy difference at small
$\Gamma$ is that $\delta f_{\rm fit}^{\rm OCP} > 0$ for $\Gamma < 17$,
which would imply that the OCP were in the solid state at very low
$\Gamma$. Note that these effects are even worse if
Eq.~(\ref{eq:fsfull}) is used to represent $f_s^{\rm OCP}$: in that
case the free energy difference decreases with $\Gamma$ for $\Gamma
\in [0,85]$ and is greater than zero for $\Gamma < 51$. To avoid
small-$\Gamma$ problems, we cut off Eq.~(\ref{eq:delfOCP}) at
$\Gamma=100$ and assume that below this value the free energy
difference is given by
\be
\delta f^{\rm OCP}(\Gamma) = -0.37+0.0046(\Gamma-100) \,,
\label{eq:delfOCPlow}
\ee
i.e., by the line tangent to $\delta f_{\rm fit}^{\rm OCP}$ at
$\Gamma=100$. If we had instead used Eq.~(\ref{eq:fsfull}) to
represent $f_s^{\rm OCP}$ in $\delta f_{\rm fit}^{\rm OCP}$,
Eq.~(\ref{eq:delfOCPlow}) would change to $\delta f^{\rm OCP}(\Gamma)
= -0.30+0.0025(\Gamma-100)$. Such a change leads to `errors' in the
multi-component results (Sections~\ref{sub:tcp} and \ref{sub:mcp}) of
no more than $5\%$ for the liquid abundances, comparable to what is
seen in Fig.~\ref{fig:delflcomp} of Appendix~\ref{sec:delfl}.

Our final expression for $\delta f^{\rm OCP}$, valid over all
$\Gamma$, is
\be
\delta f^{\rm OCP}(\Gamma) =
\begin{cases}
\delta f_{\rm fit}^{\rm OCP}(\Gamma) \,, & \hspace{-2.8em}100 < \Gamma < 200 \,, \\
-0.37+0.0046(\Gamma-100) \,, & \Gamma < 100 \,, \\
0.09+0.0043(\Gamma-200) \,, & \Gamma > 200 \,,
\end{cases}
\label{eq:delf2OCP}
\ee
where $\delta f_{\rm fit}^{\rm OCP}(\Gamma)$ is given by
Eq.~(\ref{eq:delfOCP}).

\subsection{The two-component plasma}
\label{sub:tcp}

The free energy of a two-component plasma (TCP) can be described as a
function of $N$, $T$, and the Coulomb coupling parameter $\Gamma_i =
Z_i^{5/3} \Gamma_e$ and fractional composition $x_i = N_i/N$ of either
species of ion. Here $N = N_1+N_2$ is the total number of ions and
$n_e = (Z_1 N_1 + Z_2 N_2)/V$ is the total electron density. For the
rest of this section we will identify the composition of the TCP by
$x_1$ and the Coulomb coupling parameter by $\Gamma_1$, since we can
express $x_2$ and $\Gamma_2$ as functions of these values: $x_2=1-x_1$
and $\Gamma_2=(Z_2/Z_1)^{5/3}\Gamma_1$. Note that throughout this
paper we choose to label the ionic species such that $Z_1 < Z_2 <
\cdots < Z_m$, where $m$ is the total number of species; $Z_1$ always
represents the ion with the smallest charge.

The free energy of the liquid phase of a two-component plasma is given
by
\bal
f_l^{\rm TCP}(\Gamma_1,x_1) = {}& \sum_{i=1}^2 x_i \left[f_l^{\rm OCP}(\Gamma_i) + \ln\left(x_i \frac{Z_i}{\langle Z \rangle}\right)\right] \nonumber\\
 & + \Delta f_l(\Gamma_1,x_1) \,,
\label{eq:flTCP}
\eal
where $\langle Z \rangle = \sum_{i=1}^2 x_iZ_i$ is the average ion
charge. The $\sum_{i=1}^2 x_i \ln\left(x_i \frac{Z_i}{\langle Z
\rangle}\right)$ term is the (ideal gas) entropy of mixing for two
species of volumes $Z_1 N_1/n_e$ and $Z_2 N_2/n_e$, and $\Delta f_l$
is the deviation from linear mixing in the liquid. The deviation term
$\Delta f_l$ has a similar dependence on $x_i$ to the entropy of
mixing term, but is in general much smaller in magnitude (see, e.g.,
Refs.~\cite{ogata93,dewitt96,potekhin09}). We therefore expect this
deviation to have a minimal effect on the phase transition properties
for most systems. In our calculation we set $\Delta f_l=0$ and use the
linear mixing approximation:
\be
f_l^{\rm TCP}(\Gamma_1,x_1) \simeq \sum_{i=1}^2 x_i \left[f_l^{\rm OCP}(\Gamma_i) + \ln\left(x_i \frac{Z_i}{\langle Z \rangle}\right)\right] \,.
\label{eq:flTCP2}
\ee
The error introduced by neglecting the $\Delta f_l$ term in the
expression for $f_l^{\rm TCP}$ is discussed in
Appendix~\ref{sec:delfl}.

The free energy of the solid phase of a two-component plasma is given by
\bal
f_s^{\rm TCP}(\Gamma_1,x_1) = {}& \sum_{i=1}^2 x_i \left[f_s^{\rm OCP}(\Gamma_i) + \ln\left(x_i \frac{Z_i}{\langle Z \rangle}\right)\right] \nonumber\\
 & + \Delta f_s(\Gamma_1,x_1) \,,
\label{eq:fsTCP}
\eal
where $\Delta f_s$ is the deviation from linear mixing in the
solid. Unlike $\Delta f_l$, which is generally small even at large
$\Gamma_1$ (Appendix~\ref{sec:delfl}), $\Delta f_s$ is comparable to
the other terms in $f_s$ and grows linearly with $\Gamma_1$; we
therefore expect $\Delta f_s$ to play an important role in setting the
phase transition properties. For charge ratios $R_Z = Z_2/Z_1$ in the
range $R_Z \in [1:5]$ the deviation is well fit by
\be
\Delta f_s(\Gamma_1,x_1) \simeq \Gamma_1 x_1 x_2 \Delta g(x_2,Z_2/Z_1) \,,
\label{eq:delusTCP}
\ee
where
\bal
\Delta g(x & ,R_Z) = \nonumber\\
 & \frac{C(R_Z)}{1+\frac{27(R_Z-1)}{1+0.1(R_Z-1)}\sqrt{x}(\sqrt{x}-0.3)(\sqrt{x}-0.7)(\sqrt{x}-1)} \,,
\label{eq:delvs}
\eal
\be
C(R_Z) = \frac{0.05(R_Z-1)^2}{[1+0.64(R_Z-1)][1+0.5(R_Z-1)^2]} \,.
\ee
Equation~(\ref{eq:delusTCP}) is from the Monte Carlo calculations of
\citet{ogata93}, and is accurate to within $10\%$ for $R_Z \alt 4.5$;
a similar formula (though accurate only for $R_Z \alt 2$) can be found
in \citet{dewitt03}. To estimate the error introduced to our results
by adopting Eq.~(\ref{eq:delusTCP}), we run several calculations with
a deviation of $1.1 \Delta f_s(\Gamma_1,x_1)$ and $0.9 \Delta
f_s(\Gamma_1,x_1)$ [i.e., $10\%$ higher or lower than the deviation we
use in our model]. For the TCP, we find errors in the liquid
abundances of $5\%$ or less, with the largest errors at high $\Gamma$
values and moderate charge ratios ($R_Z \sim 1.5$). For the
17-component mixture and $\Gamma$ value considered in
Section~\ref{sec:results}, the errors in the liquid abundances are
only $2\%$ or less.

\begin{figure}
\includegraphics[width=\columnwidth]{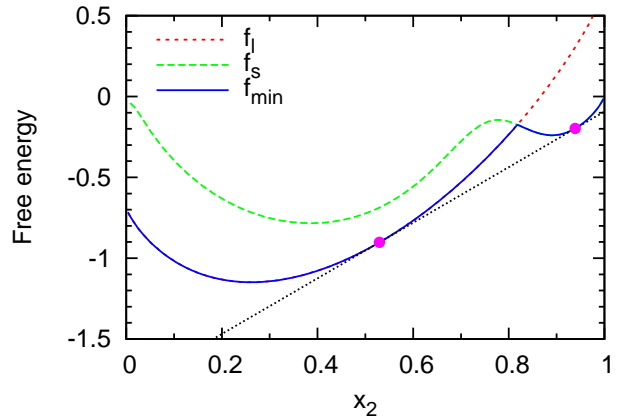}
\caption[Double-tangent construction]
{(Color online) An example of the double-tangent construction, for
$R_Z=34/8$ and $\Gamma_1=\Gamma_{\rm crit}/6$
(cf. Figs.~\ref{fig:gibbscomp} and \ref{fig:delflcomp}). The stable
compositions $a_2$ and $b_2$ (i.e., $1-a_1$ and $1-b_1$) are marked by
filled circles; here, one of the mixtures is stable in the liquid
state and one is stable in the solid state. Note that the curves $f_l$
and $f_s$ plotted in this figure are given not by
Eq.~(\ref{eq:flTCP2}) and Eq.~(\ref{eq:fsTCP}), respectively, but by
these equations minus the term $\sum_{i=1}^2 x_i f_l^{\rm
OCP}(\Gamma_i)$. The values of $a_2$ and $b_2$ obtained are the same
whether $f_l^{\rm TCP}$ and $f_s^{\rm TCP}$ or these modified
expressions are used: adding terms constant or linear in the $x_i$'s
to both free energy curves has no effect on the results of the
double-tangent construction.}
\label{fig:tangent}
\end{figure}

For a TCP at a particular value of $\Gamma_1$, we find the state of
lowest free energy as a function of composition by using the
``double-tangent'' construction (see, e.g., Ref.~\cite{gordon68}): We
construct lines tangent to the minimum free energy curve $f_{\rm
min}=\min(f_l,f_s)$ in at least two points, corresponding to the
compositions $a_1$ and $b_1$; an example of this construction is shown
graphically in Fig.~\ref{fig:tangent}. Any homogeneous composition
$x_1$ that lies between $a_1$ and $b_1$, i.e., any $x_1$ which can be
expressed as $Aa_1+(1-A)b_1 = x_1$ for some $0 < A <1$, satisfies
$Af_{\rm min}(a_1)+(1-A)f_{\rm min}(b_1) < f_{\rm min}(x_1)$ and is
therefore unstable with respect to a heterogeneous mixture of $a_1$
and $b_1$. In this paper we refer to the locus of all points
$(\Gamma_1,x_1)$ that lie between double-tangent points
$(\Gamma_1,a_1)$ and $(\Gamma_1,b_1)$ as the `unstable region' of the
phase diagram.

Note that double-tangent points $a_1$ and $b_1$ can potentially be
constructed from the liquid curve to itself, from the solid curve to
itself, or from the liquid curve to the solid curve, depending on the
behavior of $f_l$ and $f_s$ [see Eqs.~(\ref{eq:flTCP2}) and
(\ref{eq:fsTCP})]. In some cases `triple-tangent' points can be
constructed; typically this occurs when the solid curve is tangent to
itself and to the liquid curve (when the liquid is at the ``eutectic
point''; see, e.g., Ref.~\cite{segretain93}). The liquid-solid
solutions are discussed below, in Section~(\ref{sub:lsTCP}). In the
approximation we have adopted above, where the deviation from linear
mixing for the liquid is $\Delta f_l=0$, tangents to the liquid curve
$f_l$ do not intersect the curve at any other point
[cf. Eq.~(\ref{eq:flTCP2})]; therefore there are no liquid-liquid
solutions. Because of the $\Delta f_s>0$ term in the solid curve,
which grows proportional with $\Gamma_1$ [see
Eq.~(\ref{eq:delusTCP})], when $\Gamma_1$ is large enough there will
always be regions of $f_s$ where double tangents can be constructed
from the solid curve to itself. These solid-solid solutions will be
examined in a later paper.

\subsubsection{Solving for the liquid-solid equilibrium of the two-component plasma}
\label{sub:lsTCP}

For a two-component plasma, liquid-solid phase transitions occur at
compositions and $\Gamma$ values where double-tangent lines can be
drawn connecting the free energy curves of the liquid and the
solid. Under these conditions a liquid state of composition $a_1$ and
a solid state of composition $b_1$ exist simultaneously as a
mixture. For a double-tangent line connecting $f_l$ to $f_s$, the line
must satisfy
\be
f'_l(a_1) = f'_s(b_1)
\ee
and
\be
f_l(a_1) + (b_1-a_1)f'_l(a_1) = f_s(b_1) \,.
\ee
For later convenience we rewrite these equations as:
\be
f_l(a_1) + (1-a_1)f'_l(a_1) = f_s(b_1) + (1-b_1)f'_s(b_1)
\ee
and
\be
f_l(a_1) - a_1f'_l(a_1) = f_s(b_1) - b_1f'_s(b_1) \,.
\ee
Using Eqs.~(\ref{eq:flTCP2}) and (\ref{eq:fsTCP}) the system of equations to solve becomes
\bal
\delta f & {}^{\rm OCP}(\Gamma_1)+\ln\left(a_1\frac{Z_1}{\langle Z \rangle}_a\right)-\frac{Z_1}{\langle Z \rangle}_a \nonumber\\
= {}&\ln\left(b_1\frac{Z_1}{\langle Z \rangle}_b\right)-\frac{Z_1}{\langle Z \rangle}_b -\Delta f_s \nonumber\\
& + \Gamma_1b_2\left\{\Delta g(b_2,Z_2/Z_1)-b_1b_2\left[\frac{d\Delta g}{dx}\right](b_2,Z_2/Z_1)\right\} \,,
\label{eq:TCP1}
\eal
\bal
\delta f & {}^{\rm OCP}(\Gamma_2)+\ln\left(a_2\frac{Z_2}{\langle Z \rangle}_a\right)-\frac{Z_2}{\langle Z \rangle}_a \nonumber\\
= {}& \ln\left(b_2\frac{Z_2}{\langle Z \rangle}_b\right)-\frac{Z_2}{\langle Z \rangle}_b -\Delta f_s \nonumber\\
& + \Gamma_1b_1\left\{\Delta g(b_2,Z_2/Z_1)-b_1b_2\left[\frac{d\Delta g}{dx}\right](b_2,Z_2/Z_1)\right\} \,,
\label{eq:TCP2}
\eal
where $\langle Z \rangle_a = \sum a_iZ_i$, $\langle Z \rangle_b = \sum b_iZ_i$, and
\bal
& \left[\frac{d\Delta g}{dx}\right](x,R_Z) = \nonumber\\
& -\frac{C(R_Z)(2x-3\sqrt{x}+1.21-0.105/\sqrt{x})}{\left[1+\frac{27(R_Z-1)}{1+0.1(R_Z-1)}\sqrt{x}(\sqrt{x}-0.3)(\sqrt{x}-0.7)(\sqrt{x}-1)\right]^2} \,.
\label{eq:Ddelv}
\eal
[cf. Eq.~(\ref{eq:delvs})]. With these two equations (and $a_1+a_2=1$,
$b_1+b_2=1$), if we are given $\Gamma_1$ we can solve for $a_1$ and
$b_1$. This allows us to trace out the liquid-solid unstable region of
the phase diagram for $\Gamma_1$ versus $x_1$. Note that to map out
the full phase diagram we also need to know the shape of the
solid-solid unstable region; this is most important at large
$\Gamma_1$. Examples of phase diagrams for TCPs (including both types
of unstable regions) are shown in the appendices.

\subsection{The multi-component plasma}
\label{sub:mcp}

The free energy of an $m$-component plasma (MCP) can be described as a
function of $N$, $T$, the fraction composition of each ion species
$x_i = N_i/N$ (though $x_m$ is not needed, since $x_m=1-\sum x_i$),
and the Coulomb coupling parameter of one ion species. In the
following we solve for $\Gamma_1 = Z_1^{5/3} \Gamma_e$ and then use
the relation $\Gamma_i = (Z_i/Z_1)^{5/3} \Gamma_1$ to find the other
parameters.

As with the two-component plasma, the free energy of the liquid phase
of a multi-component plasma is very well described by the linear
mixing rule (but see Appendix~\ref{sec:delfl}):
\bal
f_l^{\rm MCP}(\Gamma_1,x_1, & \ldots,x_{m-1}) \nonumber\\
\simeq & \sum_{i=1}^m x_i \left[f_l^{\rm OCP}(\Gamma_i) + \ln\left(x_i \frac{Z_i}{\langle Z \rangle}\right)\right]
\label{eq:flMCP}
\eal
where $\langle Z \rangle = \sum_{i=1}^m x_iZ_i$.

The free energy of the solid phase of the MCP is
\bal
f_s^{\rm MCP}(\Gamma_1,x_1, & \ldots,x_{m-1}) \nonumber\\
 \simeq & \sum_{i=1}^m x_i \left[f_s^{\rm OCP}(\Gamma_i) + \ln\left(x_i \frac{Z_i}{\langle Z \rangle}\right)\right] \nonumber\\
 & + \Delta f_s(\Gamma_1,x_1,\ldots,x_{m-1}) \,.
\label{eq:fsMCP}
\eal
According to \citet{ogata93}, the deviation of the solid from linear
mixing $\Delta f_s$ for a three-component plasma is given to good
accuracy by
\bal
\Delta f_s(\Gamma_1,x_1, & \ldots,x_{m-1}) \nonumber\\
\simeq & \sum_{i=1}^{m-1}\sum_{j=i+1}^m \Gamma_i x_i x_j \Delta g\left(\frac{x_j}{x_i+x_j},\frac{Z_j}{Z_i}\right) \,,
\label{eq:delusMCP}
\eal
where $Z_1 < Z_2 < \cdots < Z_m$ and $\Delta g(x,R_Z)$ is given by
Eq.~(\ref{eq:delvs}). We assume here that Eq.~(\ref{eq:delusMCP})
applies for all $m \ge 2$. A partial justification for this assumption
is provided in Appendix~\ref{sec:delfs}.

In the $m$-component plasma we construct $(m-1)$-dimensional
hyperplanes tangent to the minimum free energy surface in at least two
points, corresponding to the compositions $\vec{a}$ and $\vec{b}$. Any
homogeneous composition $\vec{x}$ that lies between $\vec{a}$ and
$\vec{b}$, i.e., any $\vec{x}$ which can be expressed as
$A\vec{a}+(1-A)\vec{b} = \vec{x}$ for some $0 < A < 1$, is unstable
with respect to a heterogeneous mixture of $\vec{a}$ and $\vec{b}$.

\subsubsection{Solving for the liquid-solid equilibrium of the multi-component plasma}
\label{sub:lsMCP}

For a multi-component plasma, liquid-solid phase transitions occur at
compositions and $\Gamma$ values where double-tangent hyperplanes can be
drawn connecting the free energy surfaces of the liquid and the
solid. For a double-tangent hyperplane connecting $f_l(\vec{a})$ to
$f_s(\vec{b})$, the hyperplane must satisfy
\be
\frac{df_l}{dx_i}(\vec{a}) = \frac{df_s}{dx_i}(\vec{b}) \,, \qquad i \in [1,m-1]
\ee
and
\be
f_l(\vec{a}) + \left(\vec{b}-\vec{a}\right) \cdot \nabla f_l(\vec{a}) = f_s(\vec{b}) \,;
\ee
or
\bal
f_l(\vec{a}) + \frac{df_l}{dx_i}(\vec{a}) - \vec{a} \cdot \nabla f_l(\vec{a}) & = f_s(\vec{b}) + \frac{df_s}{dx_i}(\vec{b}) - \vec{b} \cdot \nabla f_s(\vec{b}) \,, \nonumber\\
i & \in [1,m-1]
\eal
and
\be
f_l(\vec{a}) - \vec{a} \cdot \nabla f_l(\vec{a}) = f_s(\vec{b}) - \vec{b} \cdot \nabla f_s(\vec{b})
\ee

Using Eqs.~(\ref{eq:flMCP}) and (\ref{eq:fsMCP}) the system of equations to solve becomes
\begin{widetext}
\bal
\delta f^{\rm OCP}(\Gamma_i)+\ln\left(a_i\frac{Z_i}{\langle Z \rangle}_a\right)-\frac{Z_i}{\langle Z \rangle}_a
= {}& \ln\left(b_i\frac{Z_i}{\langle Z \rangle}_b\right)-\frac{Z_i}{\langle Z \rangle}_b -\Delta f_s(\Gamma_1,x_1,\ldots,x_{m-1}) \nonumber\\
 & + \sum_{j=1}^{i-1} \Gamma_j b_j\left\{\Delta g\left(\frac{b_i}{b_i+b_j},\frac{Z_i}{Z_j}\right)+\frac{b_ib_j}{(b_i+b_j)^2}\left[\frac{d\Delta g}{dx}\right]\left(\frac{b_i}{b_i+b_j},\frac{Z_i}{Z_j}\right)\right\} \nonumber\\
 & + \sum_{j=i+1}^{m} \Gamma_i b_j\left\{\Delta g\left(\frac{b_j}{b_i+b_j},\frac{Z_j}{Z_i}\right)-\frac{b_ib_j}{(b_i+b_j)^2}\left[\frac{d\Delta g}{dx}\right]\left(\frac{b_j}{b_i+b_j},\frac{Z_j}{Z_i}\right)\right\} \,,
\label{eq:MCPn}
\eal
\end{widetext}
for $i \in [1,m]$. Here $\langle Z \rangle_a = \sum a_iZ_i$, $\langle
Z \rangle_b = \sum b_iZ_i$, and $\left[\frac{d\Delta
v}{dx}\right](x,R_Z)$ is again given by Eq.~(\ref{eq:Ddelv}). With
these $m$ equations (and $\sum a_i=1$, $\sum b_i=1$), if we are given
the liquid composition $\vec{a}$ we can solve for the solid
composition $\vec{b}$ and Coulomb parameter $\Gamma_1$ at which the
liquid and solid states are in equilibrium; if we are given $\vec{b}$
we can solve for $\vec{a}$ and $\Gamma_1$. In this manner we can trace
out the liquid-solid unstable region of the phase diagram for
$\Gamma_1$ versus $\vec{x}$. As in the TCP case, to map out
the full phase diagram we also need to know the shape of the the
solid-solid unstable region.

Alternatively, if we are given an initial composition $\vec{x}$ and
the fraction $0<A<1$ of the solution in the liquid state (or the
fraction $1-A$ in the solid state), we can solve for $\Gamma_1$ and
the compositions of both the liquid and solid mixtures in
equilibrium. We have $2m-1$ unknowns, $a_1,\ldots,a_{m-1}$,
$b_1,\ldots,b_{m-1}$, and $\Gamma_1$; but in addition to the the $m$
equations Eq.~(\ref{eq:MCPn}) above we have the $m-1$ equations
\be
Aa_i+(1-A)b_i = x_i \,, \qquad i \in [1,m-1] \,.
\label{eq:Aeq}
\ee

\section{Results}
\label{sec:results}

As described in Section~\ref{sec:intro}, \citet{horowitz07} [hereafter
HBB] use a molecular dynamics simulation to study the phase transition
of a 17-component plasma. A total of $27,648$ ions are placed in a
simulation volume of length $727.5$~fm on a side. At the start of the
simulation $50\%$ of the plasma is in the liquid state and $50\%$ is
in the solid state. There is a uniform composition throughout the
volume, given by the results of \citet{gupta07} (who calculate the
composition of an accreting neutron star at a density of
$2\times10^{11}$~g/cm$^3$, after the accreted material has undergone
proton and electron capture and various other reactions). As the
system evolves, the temperature is adjusted so that approximately half
of the plasma remains in the liquid state and half remains in the
solid state. After a simulation time of $5\times10^6$~fm$/c$, the
simulation is run at constant energy until the total time reaches
$151\times 10^6$~fm$/c$. The results of the numerical simulation are
shown in Table~\ref{tab:HBB}. The final temperature of the simulation
is expressed in terms of $\Gamma_1$ as well as the `average' Coulomb
coupling parameter, $\Gamma = \langle Z^{5/3} \rangle \Gamma_e$. For
each entry in Table~\ref{tab:HBB}, a statistical ($\sqrt{N_i}$) error
is provided.

\begin{table*}
\caption{Abundance of chemical element $Z$, for various mixtures from
the numerical simulation of \citet{horowitz07}. Abundances are
provided for the initial mixture (in the column labeled `Initial') and
the final liquid and solid mixtures (in the columns labeled ``Liquid''
and ``Solid'', respectively). For each final mixture, the average charge
$\langle Z \rangle$ and Coulomb coupling parameter $\Gamma = \langle
Z^{5/3} \rangle \Gamma_e$ are provided as well. The percentage error
for each entry is given by $100/\sqrt{N_i}$, where $N_i = x_i N$ and
$N=27,648$.}
\centering
\begin{tabularx}{0.75\textwidth}{r Y Y@{\quad}r Y@{\quad}r}
\hline\hline
\multicolumn{6}{c}{HBB results} \\
\multicolumn{6}{c}{$\langle Z \rangle_l=28.04$, $\langle Z \rangle_s=30.48$} \\
\multicolumn{6}{c}{$\Gamma_1=27.7$, $\Gamma_l=233$, $\Gamma_s=261$} \\
Z\, & Initial\, & Liquid & $\%$ Error & Solid\,\, & $\%$ Error \\
\hline
 8 & 0.0301 & 0.0529 &  3 & 0.0087 &  6 \\
10 & 0.0116 & 0.0205 &  4 & 0.0021 & 13 \\
12 & 0.0023 & 0.0043 &  9 & 0.0006 & 24 \\
14 & 0.0023 & 0.0043 &  9 & 0.0005 & 27 \\
15 & 0.0023 & 0.0043 &  9 & 0.0004 & 30 \\
20 & 0.0046 & 0.0055 &  8 & 0.0029 & 11 \\
22 & 0.0810 & 0.1024 &  2 & 0.0616 &  2 \\
24 & 0.0718 & 0.0816 &  2 & 0.0635 &  2 \\
26 & 0.1019 & 0.1065 &  2 & 0.1017 &  2 \\
27 & 0.0023 & 0.0025 & 12 & 0.0027 & 12 \\
28 & 0.0764 & 0.0744 &  2 & 0.0746 &  2 \\
30 & 0.0856 & 0.0773 &  2 & 0.0949 & 20 \\
32 & 0.0116 & 0.0099 &  6 & 0.0130 &  5 \\
33 & 0.1250 & 0.1079 &  2 & 0.1388 &  1 \\
34 & 0.3866 & 0.3408 &  1 & 0.4297 & 0.9 \\
36 & 0.0023 & 0.0012 & 17 & 0.0030 & 11 \\
47 & 0.0023 & 0.0030 & 11 & 0.0013 & 17 \\
\hline\hline
\end{tabularx}
\label{tab:HBB}
\end{table*}

We have applied our semi-analytic calculation (Section~\ref{sub:mcp})
to the same 17-component mixture as is considered by HBB. In
Eq.~(\ref{eq:Aeq}) we set $\vec{x}$ to the `initial' composition given
in Table~\ref{tab:HBB}, and choose $A=0.5$, such that we are solving
for the equilibrium state where $50\%$ of the mixture is liquid and
$50\%$ is solid. We then use Eqs.~(\ref{eq:MCPn}) and (\ref{eq:Aeq})
to find the final composition of the liquid and solid states,
$\vec{a}$ and $\vec{b}$. The result is given in
Table~\ref{tab:instant}. For each entry in Table~\ref{tab:instant}, an
error is provided in terms of the percent difference from the
corresponding HBB result.

The results of Table~\ref{tab:instant} are relevant under equilibrium
conditions, which in the accreting neutron star means that the
particles solidify and diffuse through the liquid and the solid faster
than new material is accreted. Here we attempt to estimate the
importance of the diffusion rate on the overall results. In order to
do that, we repeat our calculation done with `instantaneous diffusion'
(Table~\ref{tab:instant}), this time assuming `no diffusion' in the
solid \footnote{In both calculations we assume that the liquid
diffusion is rapid. This is usually the case for terrestrial mixtures
undergoing phase transitions (e.g., Ref.~\cite{gordon68}), but it
appears to be true for the ocean of an accreting neutron star as well
(C. Horowitz, private communication).}. As in the equilibrium case, the
calculation starts with the plasma in the liquid state with initial
composition given by HBB, and ends when $50\%$ of the plasma is liquid
and $50\%$ is solid. Unlike in the equilibrium case, however, we solve
Eqs.~(\ref{eq:MCPn}) and (\ref{eq:Aeq}) many times, each time
producing a small amount of solid material ($1-A \ll 1$). Solid
particles created in one step are removed from consideration in all
future steps, since we are assuming that these particles do not mix.
The liquid composition ($\vec{a}$) calculated in one step is used as
the `initial' composition ($\vec{x}$) in the next step.

While an exact treatment of the `no diffusion' limit would require
solving Eqs.~(\ref{eq:MCPn}) and (\ref{eq:Aeq}) on a
particle-to-particle basis, we find that a good approximation can be
obtained using $500$ steps with $A_k = 1-1/(1001-k)$ for each step
$k$. [The difference between the final abundances calculated using $50$
steps with $A_k = 1-1/(101-k)$ and $500$ steps with $A_k = 1-1/(1001-k)$,
e.g., is less than $0.2\%$.] The result is given in
Table~\ref{tab:nodiff}. Note that for this choice for $A_k$, the
number of solid particles created is the same in each step. The
average solid composition is given by
\be
\vec{\langle b\rangle} = \frac{1}{50}\sum_{k=1}^{50} \vec{b}^k \,,
\ee
where $\vec{b}^k$ is the composition of the solid particles created in
the $k$th step.

A comparison of Tables~\ref{tab:instant} and \ref{tab:nodiff} shows
that calculations done under the two diffusion limits give very
similar results. For example, the abundance differences between these
two calculations are generally much smaller than between either
calculation and the results of HBB. Therefore, we conclude that the
error introduced into our calculation by assuming instantaneous
diffusion rather than the actual diffusion rate (whatever that may be)
is small. Note that even though the rate of diffusion has very little
effect on the average composition in the solid, it has a strong effect
on the how that composition varies locally. For sufficiently low
diffusion rates, lamellar sheets or other structures may form in the
solid (see, e.g., Ref.~\cite{gordon68}); these structures can have a
strong effect on the thermal conductivity and strength of the crust.

\begin{table*}
\caption{Abundance of chemical element $Z$, for the liquid and solid
mixtures from our equilibrium calculation. Here, instantaneous
diffusion is assumed (see text). For each mixture, the average charge
$\langle Z \rangle$ and Coulomb coupling parameter $\Gamma = \langle
Z^{5/3} \rangle \Gamma_e$ are provided as well. The initial liquid
mixture is given by its value from HBB, and the system is evolved
until there is $50\%$ liquid material, $50\%$ solid material. The
percent error for each entry is given by $100 \times ({\rm entry}-{\rm
HBB})/{\rm HBB}$.}
\centering
\begin{tabularx}{0.75\textwidth}{r Y Y@{\quad}r Y@{\quad}r}
\hline\hline
\multicolumn{6}{c}{Instant diffusion} \\
\multicolumn{6}{c}{$\langle Z \rangle_l=27.667$, $\langle Z \rangle_s=30.930$} \\
\multicolumn{6}{c}{$\Gamma_1=26.57$, $\Gamma_l=218.3$, $\Gamma_s=256.1$ ($\Gamma_1$ error: $-4\%$)} \\
Z & Initial\, & Liquid & $\%$ Error & Solid\,\, & $\%$ Error \\
\hline
 8 & 0.0301 & 0.0513 & -3 & 0.0089 & +3 \\
10 & 0.0116 & 0.0197 & -4 & 0.0035 & +66 \\
12 & 0.0023 & 0.0039 & -8 & 0.0007 & +10 \\
14 & 0.0023 & 0.0040 & -7 & 0.0006 & +22 \\
15 & 0.0023 & 0.0040 & -7 & 0.0006 & +54 \\
20 & 0.0046 & 0.0073 & +32 & 0.0019 & -33 \\
22 & 0.0810 & 0.1213 & +18 & 0.0407 & -34 \\
24 & 0.0718 & 0.0947 & +16 & 0.0489 & -23 \\
26 & 0.1019 & 0.1161 & +9 & 0.0877 & -14 \\
27 & 0.0023 & 0.0024 & -4 & 0.0022 & -19 \\
28 & 0.0764 & 0.0758 & +2 & 0.0770 & +3 \\
30 & 0.0856 & 0.0759 & -2 & 0.0953 & +0.5 \\
32 & 0.0116 & 0.0095 & -4 & 0.0137 & +5 \\
33 & 0.1250 & 0.1013 & -6 & 0.1487 & +7 \\
34 & 0.3866 & 0.3076 & -10 & 0.4656 & +8 \\
36 & 0.0023 & 0.0018 & -12 & 0.0028 & -8 \\
47 & 0.0023 & 0.0033 & +9 & 0.0013 & +2 \\
\hline\hline
\end{tabularx}
\label{tab:instant}
\end{table*}

\begin{table*}
\caption{As in Table~\ref{tab:instant}, except that diffusion is
assumed to be negligible in the solid (see text).}
\centering
\begin{tabularx}{0.75\textwidth}{r Y Y@{\quad}r Y@{\quad}r}
\hline\hline
\multicolumn{6}{c}{No diffusion} \\
\multicolumn{6}{c}{$\langle Z \rangle_l=27.370$, $\langle Z \rangle_s=30.680$} \\
\multicolumn{6}{c}{$\Gamma_1=27.38$, $\Gamma_l=221.2$, $\Gamma_s=260.6$ ($\Gamma_1$ error: $-1\%$)} \\
Z & Initial\, & Liquid & $\%$ Error & Solid\,\, & $\%$ Error \\
\hline
 8 & 0.0301 & 0.0526 & -0.5 & 0.0076 & -13 \\
10 & 0.0116 & 0.0204 & -0.5 & 0.0028 & +34 \\
12 & 0.0023 & 0.0041 & -5 & 0.0005 & -14 \\
14 & 0.0023 & 0.0041 & -4 & 0.0005 & -5 \\
15 & 0.0023 & 0.0041 & -4 & 0.0005 & +19 \\
20 & 0.0046 & 0.0077 & +39 & 0.0015 & -47 \\
22 & 0.0810 & 0.1289 & +26 & 0.0331 & -46 \\
24 & 0.0718 & 0.1018 & +25 & 0.0418 & -34 \\
26 & 0.1019 & 0.1240 & +16 & 0.0798 & -22 \\
27 & 0.0023 & 0.0025 & +1 & 0.0021 & -23 \\
28 & 0.0764 & 0.0786 & +6 & 0.0742 & -0.5 \\
30 & 0.0856 & 0.0753 & -3 & 0.0959 & +1 \\
32 & 0.0116 & 0.0091 & -8 & 0.0141 & +8 \\
33 & 0.1250 & 0.0953 & -12 & 0.1547 & +11 \\
34 & 0.3866 & 0.2863 & -16 & 0.4869 & +13 \\
36 & 0.0023 & 0.0017 & -18 & 0.0029 & -4 \\
47 & 0.0023 & 0.0034 & +15 & 0.0012 & -11 \\
\hline\hline
\end{tabularx}
\label{tab:nodiff}
\end{table*}

A comparison of Tables~\ref{tab:instant} and \ref{tab:nodiff} to
Table~\ref{tab:HBB} shows that the semi-analytic calculation does
quite well at reproducing the results of the HBB numerical
simulation. All of the abundances from the semi-analytic calculation
are with $65\%$ of the HBB values, and most are significantly closer.
Also, many of the table entries that match poorly between the two
works correspond to chemical elements with very low abundances, i.e.,
those elements that are most affected by the finite size of the
simulation. For example, the two entries that match the worst between
Tables~\ref{tab:HBB} and \ref{tab:instant}, the solid abundances of
elements $Z=10$ and $15$, are represented in the simulation by only
$58$ and $11$ ions, respectively.

\begin{figure}
\includegraphics[width=\columnwidth]{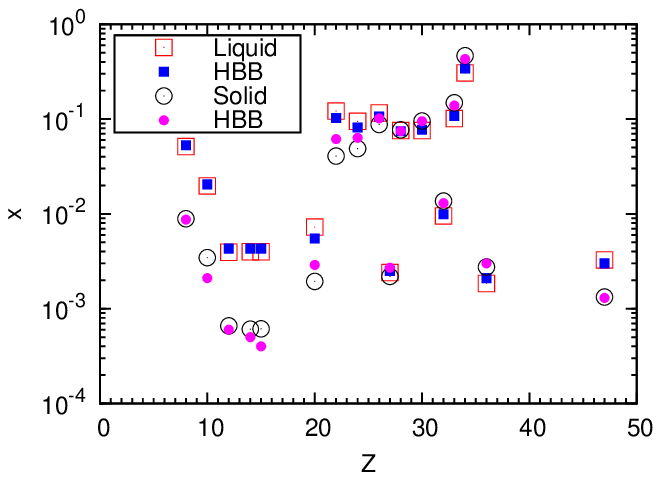}
\caption[Abundances of the liquid and solid mixtures]
{(Color online) Abundances $x$ as a function of chemical element $Z$,
for the final liquid and solid mixtures. Both the values from our
equilibrium calculation (``Liquid'' and ``Solid'', the large open
squares and circles, respectively) and from the numerical simulation
of HBB (``HBB'', the small filled squares and circles) are shown.}
\label{fig:absolute}
\end{figure}

\begin{figure}
\includegraphics[width=\columnwidth]{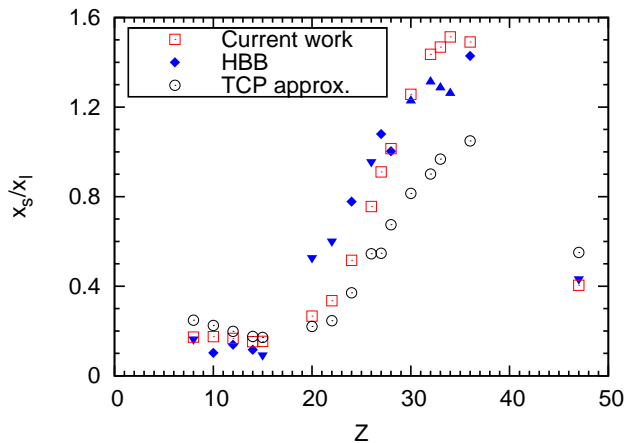}
\caption[Ratio of the liquid to solid abundances]
{(Color online) Ratio of the solid abundance to the liquid abundance
$x_s/x_l$ as a function of chemical element $Z$. Both the values from
our equilibrium calculation (``Current work'', the open squares) and
from the numerical simulation of HBB (``HBB'', the filled diamonds and
triangles) are shown, as are the values predicted from the
`two-component' approximation (``TCP approx.'', the open circles; see
text). If for a given element the HBB ratio is still evolving at the
end of the simulation, it is plotted with a triangle that points in
the direction of evolution; if the ratio is not changing or is
oscillating up and down, it is plotted with a diamond.}
\label{fig:ratio}
\end{figure}

Figures~\ref{fig:absolute} and \ref{fig:ratio} provide further
comparison of our results and those of HBB. Figure~\ref{fig:absolute}
(cf. Fig.~2 of HBB) presents in graphical form the data from
Tables~\ref{tab:HBB} and \ref{tab:instant}, i.e., the final
compositions of the liquid and solid states for both the HBB numerical
simulation and our semi-analytic calculation. Figure~\ref{fig:ratio}
(cf. Fig.~6 of HBB) shows the ratio of the solid abundance to the
liquid abundance versus atomic number $Z$ for both works.

Also plotted in Fig.~\ref{fig:ratio} are the abundance ratios in the
`two-component' approximation. In this approximation, the abundance
ratios for each element are calculated assuming the plasma is composed
of only two ion species, the element itself and the most abundant
element in the mixture (i.e., $i=15$ or $Z=34$; see
Table~\ref{tab:HBB}). The initial composition of the mixture is chosen
such that the ratio of the abundances of the two elements is the same
as in HBB (e.g., $x_1/x_{15} = 0.0301/0.3866$, but now
$x_1+x_{15}=1$); however, the results do not change much qualitatively
if we choose some other scheme. As with the 17-component plasma, we
solve for the point where half of the plasma is liquid and half is
solid. Note that the $Z=34$ abundance ratio is not plotted in
Fig.~\ref{fig:ratio} for this approximation, as its value is different
for each two-element pairing. The two-component approximation
reproduces the abundance ratio trend of the 17-component plasma,
including the relatively constant behavior at low $Z$ and the peak at
$Z=34$. It does not give accurate absolute values of the ratios,
particularly for $Z$ around $Z=34$ (where the true solid-to-liquid
ratio is greater than unity).

The abundances listed in Table~\ref{tab:HBB} are the compositions of
the HBB liquid and solid states at the end of the simulation. These
results may not represent the true equilibrium state of the mixture
because of the finite run time of the simulation. To show this effect,
the HBB abundance ratios are plotted in Fig.~\ref{fig:ratio} using one
of three symbols: for a given chemical element, if at the end of the
simulation run the ratio is evolving upward in time, it is plotted
with an upward-pointing triangle; if the ratio is evolving downward in
time, it is plotted with a downward-pointing triangle; and if the
ratio is not changing or is oscillating upward and downward, it is
plotted with a diamond. The determination of the evolution direction
for each element is made using data from the simulation time steps
$t_6 = t/(10^6~{\rm fm}/c) = 71$, $113$, and $151$, i.e., the last
three time steps shown in Fig.~6 of HBB. If the abundance ratio
decreases (increases) from $t_6=71$ to $113$ and from $t_6=113$ to
$151$, and the total decrease (increase) across both time intervals is
more than 0.1, the ratio is said to be evolving downward (upward) in
time; otherwise the ratio is said to be stable. Note that, for the
most part, the HBB results are evolving toward the equilibrium values
found in our calculation; this behavior is especially apparent for $Z
\in [20,34]$, which is also where the abundance ratios differ in the
two works by their largest values \footnote{After their work was
published, HBB ran their simulation an additional $208\times
10^6$~fm$/c$, to a total simulation time of $359\times
10^6$~fm$/c$. Of the solid-to-liquid abundance ratios that were still
evolving at the time of the HBB publication (i.e., those presented
with upward- or downward-pointing triangles in our
Fig.~\ref{fig:ratio}), by $359\times 10^6$~fm$/c$ just over half had
evolved closer to our results ($Z=15$, $20$, $22$, $30$, $32$, and
$34$), while the rest either had remained steady ($Z=8$) or had
evolved farther away ($Z=26$, $33$, and $47$) [D. Berry, private
communication].}. This suggests that the errors given in
Tables~\ref{tab:instant} and \ref{tab:nodiff} are strong upper limits
to the actual accuracy of our calculation.

\section{Discussion}
\label{sec:discuss}

Using results from simulations of one-, two-, and three-component
plasmas, we have developed a method for calculating the equilibrium
properties of the liquid-solid phase transition in a plasma with an
arbitrary number of components, in the approximation of a classical
ion plasma in a uniform electron background. We used this method to
calculate the phase transition properties for a 17-component plasma
with a composition similar to that which might exist in the ocean of
an accreting neutron star, and compared the results to those of a
molecular dynamics simulation done at the same composition (HBB
\cite{horowitz07}). We found that our method accurately reproduces the
results of the HBB simulation. Two sources of error in the simulation
may mean that our results represent the actual system even more
accurately than this comparison suggests: First, the finite size of
the simulation introduces statistical errors which for some components
are larger than the discrepancies between the two works. Second, the
system is still evolving at the end of the simulation, with many
components approaching the values predicted by our calculation.

As in the simulation of HBB, we have followed the 17-component mixture
until it reaches the state of $50\%$ liquid and $50\%$ solid. Under
these conditions, the term representing the deviation from the linear
mixing rule for the solid, $\Delta f_s$, is a perturbation on the
other terms in the free energy of the solid [see
Eq.~(\ref{eq:fsMCP})]. In principle our calculation can continue to
larger fractions of solid, i.e., larger values of the Coulomb coupling
parameter $\Gamma$. However, because $\Delta f_s$ increases linearly
with $\Gamma$ and eventually dominates the free energy, the
calculation at $\Gamma$ above the half-freezing point is more
sensitive to the form chosen for $\Delta f_s$. There is some numerical
confirmation of our simple approximation for $\Delta f_s$,
Eqs.~(\ref{eq:delvs}) and (\ref{eq:delusMCP}), for two- and
three-component mixtures at large $\Gamma$, but only for a very
limited set of parameters (see Ref.~\cite{ogata93}). Further numerical
simulations are necessary to test the validity of these equations at
large $\Gamma$ for general parameters and $(m>3)$-component plasmas.

Another consequence of the large and positive $\Delta f_s$ term is
that for certain compositions, it is energetically favorable for a
single solid phase to separate into two or more solid phases (see
Section~\ref{sub:tcp}). Such a phase separation occurs at large
$\Gamma$ in the 17-component plasma simulated by \citet{horowitz09a}.
With our calculation we have not yet found any two-solid mixtures that
represent the lowest energy state of the HBB plasma, in part because
the shape of free energy surface for the solid phase is very
complicated at large $\Gamma$. We leave a more careful study of the
solid-solid unstable region for future work.

Once these issues are resolved, our calculation will allow the
complete phase diagram of multi-component mixtures to be determined.
We expect that these results will have important implications for the
structure of the liquid-solid boundary in accreting neutron stars. For
example, for an ocean temperature of $T=10^8 T_8~{\rm K}$, an O-Se
mixture with the same proportion of oxygen and selenium as in the HBB
mixture (i.e., $\sim 10\%$-$90\%$) will begin to freeze at a density
of $\rho \simeq 2\times 10^7 T_8^3 (\mu_e/2)~{\rm g/cm^3}$, where
$\mu_e$ is the mean molecular weight per electron. Assuming that
accretion is slow enough that the liquid and solid can come into
equilibrium at each depth, our phase diagram for a charge ratio
$R_Z=34/8$ in Fig.~\ref{fig:gibbscomp} (or Fig.~\ref{fig:delflcomp})
shows that the mixture will reach $50\%$ solid within a factor of two
in density, but that complete freezing will not occur until much
deeper, by a factor of $\simeq (34/8)^5 \simeq 1400$ in density
(corresponding to $\rho \simeq 3\times 10^{10} T_8^3~{\rm
g/cm^3}$). This is a very different picture than the sharp transition
between liquid and solid expected for a one-component plasma, and
assumed in previous work on accreting neutron stars. Further work is
needed to understand the effects of the various time-dependent
processes that are active concurrent with accretion in the ocean-crust
transition layer, such as crystallization, diffusion, and
sedimentation. For example, sedimentation of the heavier solid
particles could be important at low accretion rates, narrowing the
transition layer.

\section*{Acknowledgments}

We thank Charles Horowitz, Alexander Potekhin, and Don Berry for
helpful discussions. ZM has been supported in part by the Lorne
Trottier Chair in Astrophysics and Cosmology at McGill University. AC
acknowledges support from an NSERC Discovery Grant and the Canadian
Institute for Advanced Research (CIFAR).

\appendix

\section{The Helmholtz free energy versus the Gibbs free energy}
\label{sec:gibbs}

Because phase transitions in stars occur at constant pressure, not
constant volume, the energy which is at a minimum when the system is
in equilibrium is the Gibbs free energy, i.e., $G = F+PV$. We discuss
here how our results (Section~\ref{sec:results}) change when the Gibbs
free energy, rather than the Helmholtz free energy, is used to
determine the equilibrium state.

To calculate the Gibbs free energy, we follow the perturbation method
of \citet{ogata93}, though we ignore terms due to the electron
exchange energy (see, e.g., Ref.~\cite{salpeter61}; these terms are
small for highly-relativistic plasmas such as are found at the
ocean-crust boundaries of accreting neutron stars). In the degenerate
interiors of white dwarfs and neutron stars, the electrons make the
dominant contribution to the total pressure ($P_i \sim \alpha \langle
Z^{2/3} \rangle P_e$ for $\Gamma>1$; see, e.g.,
Ref.~\cite{haensel07}), and so we can treat the ion partial pressures
as perturbations.

The Helmholtz free energy of the system is
\be
F = F_0 + F_1 \,,
\ee
where $F_0$ is the kinetic energy of the electrons and $F_1$ is the
free energy of the ions (the electron exchange term is ignored and the
Coulomb term is folded into the ion free energy). The total pressure
of the system is
\be
P \equiv -\frac{\partial F}{\partial V} = -\frac{\partial F_0}{\partial V} - \frac{\partial F_1}{\partial V} \,.
\label{eq:press}
\ee
Let $V_0$ be the volume of the unperturbed system, when only electrons
contribute to the total pressure; let $V_{01}$ be the volume of the
perturbed system, when both ions and electrons contribute to the total
pressure. Then the total pressure can also be expressed as
\be
P = -F'_0(V_0)
\label{eq:epress}
\ee
and
\be
P = P(V_0) + \delta V P'(V_0) + \frac{\delta V^2}{2} P''(V_0) + \cdots \,,
\label{eq:taylorpress}
\ee
where $\delta V = V_{01}-V_0$ and we are using the notation $P'(V_0) =
\left[\frac{\partial P}{\partial V}\right]_{V=V_0}$, etc. From
Eqs.~(\ref{eq:press})--(\ref{eq:taylorpress}), and assuming $\delta V$
is small (which can easily be checked {\it a posteriori}), we obtain
\be
\delta V = -\frac{F'_1(V_0)}{F''_0(V_0)} \,.
\label{eq:deltaV}
\ee
The Gibbs free energy can be written as
\bal
G = {}& G(V_0) + \delta V G'(V_0) + \frac{\delta V^2}{2} G''(V_0) + \cdots
\label{eq:taylorgibbsA} \\
 = {}& F_0(V_0)+F_1(V_0) + P(V_0)V_0 + \delta V P'(V_0) V_0 \nonumber\\
 & + \frac{\delta V^2}{2} P''(V_0) V_0 + \frac{\delta V^2}{2} P'(V_0) + \cdots
\label{eq:taylorgibbsB} \\
 = {}& F_0(V_0)+F_1(V_0) + PV_0 - \frac{\left[F'_1(V_0)\right]^2}{2 F''_0(V_0)} \,,
\label{eq:taylorgibbsC}
\eal
where in going from Eq.~(\ref{eq:taylorgibbsA}) to
Eq.~(\ref{eq:taylorgibbsB}) we have made use of the thermodynamic
relation
\be
V \equiv -\frac{\partial G}{\partial P} \,.
\ee

The Gibbs free energy is obtained from Eq.~(\ref{eq:taylorgibbsC}),
once the value of $V_0$ is known. For a given total pressure $P$, the
volume $V_0$ is determined by Eq.~(\ref{eq:epress}): We have (e.g.,
Ref.~\cite{salpeter61})
\bal
P = {}& -F'_0(V_0) \nonumber\\
 = {}& \frac{m_ec^2}{8\pi^2 \lambdabar_c^3} \left[y\sqrt{1+y^2}\left(\frac{2y^2}{3}-1\right) + \ln \left(y+\sqrt{1+y^2}\right)\right] \,,
\label{eq:epress2}
\eal
where the ``relativity parameter''
\be
y \equiv \frac{p_F}{m_ec} = \lambdabar_c(3\pi^2 n_e)^{1/3} = \left(\frac{9\pi}{4}\right)^{1/3}\frac{k_BT}{\alpha m_ec^2} \Gamma_e
\label{eq:yparam}
\ee
is evaluated at $V=V_0$. Here $\alpha = e^2/(\hbar c)$ is the fine
structure constant and $\lambdabar_c = \hbar/(m_ec)$ is the reduced
Compton wavelength. The volume $V_0$ depends only on the total
pressure of the system, and so is the same for both the liquid and
solid states. The Helmholtz free energy in the unperturbed state,
$F_0(V_0)$, is also the same for both states. We can therefore ignore
the $F_0(V_0)$ and $PV_0$ terms in Eq.~(\ref{eq:taylorgibbsC}) when
calculating the state of lowest free energy. Using
\be
F''_0(V_0) = \frac{1}{V_0}\frac{m_ec^2}{9\pi^2\lambdabar_c^3}\frac{y^5}{\sqrt{1+y^2}} \,,
\ee
we arrive at our final expression for the Gibbs free
energy of the liquid ($i=l$) or solid ($i=s$) state:
\bal
g_i \equiv {}& \frac{G_i}{Nk_BT} \nonumber\\
= {}& f_i(\Gamma_e) - \frac{\alpha}{3(18\pi)^{1/3}\langle Z \rangle}\frac{\sqrt{1+y^2}}{y} \Gamma_e\left[\frac{\partial f_i}{\partial \Gamma_e}\right]^2 \,,
\label{eq:gibbsion}
\eal
where $f_i$ is the Helmholtz free energy given in
Sections~\ref{sub:ocp}--\ref{sub:mcp}, $y(P)$ is found from
Eq.~(\ref{eq:epress2}), and $\Gamma_e(y)$ is found from
Eq.~(\ref{eq:yparam}) (i.e., $\Gamma_e$ is evaluated at $V=V_0$).

\begin{figure}
\begin{center}
\includegraphics[width=\columnwidth]{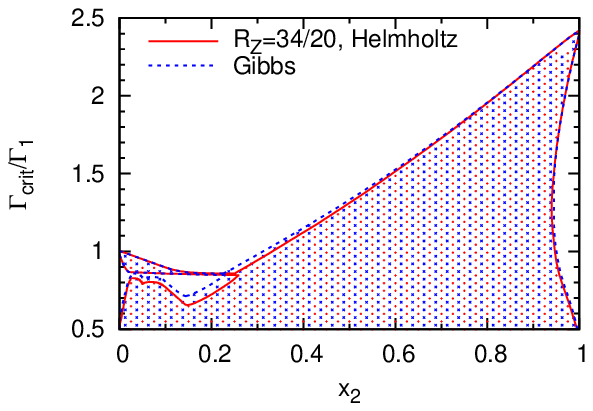}
\includegraphics[width=\columnwidth]{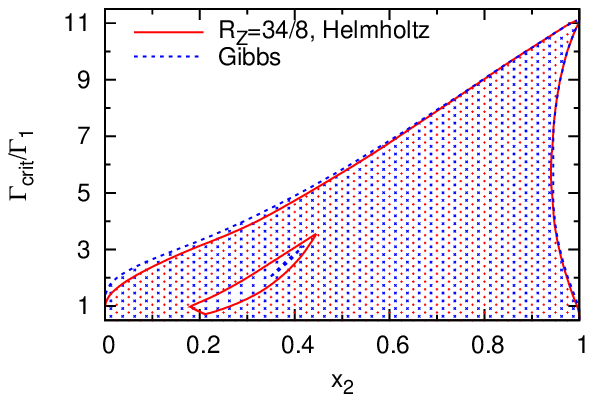}
\end{center}
\caption[Phase diagrams at constant volume and at constant pressure]
{(Color online) Phase diagrams for charge ratios $R_Z = 34/20$ (top
panel) and $R_Z = 34/8$ (bottom panel). Phase transitions at constant
volume are labeled ``Helmholtz'', and transitions at constant pressure
are labeled ``Gibbs''. To maintain consistency with earlier works
(e.g., Refs.~\cite{segretain93,ogata93}), $\Gamma_1^{-1}$ in units of
$\Gamma_{\rm crit}^{-1}$ is plotted versus $x_2$, where $Z_2=34$ for
all transitions. The unstable regions are marked by dots. The mixture
is liquid for $(x_2,\Gamma_{\rm crit}/\Gamma_1)$ points entirely above
the unstable region; for points below any part of the unstable region
(such as the peninsula in the bottom-left corner of the top panel and
the banana-shaped island in the bottom panel) the mixture is solid.}
\label{fig:gibbscomp}
\end{figure}

We calculate the phase diagrams for two-component plasmas with charge
ratios $R_Z=Z_2/Z_1$ up to $34/8$, first using the relevant
expressions for $f_l$ and $f_s$ from Section~\ref{sub:tcp} (i.e.,
ignoring pressure terms), and then using Eq.~(\ref{eq:gibbsion})
(including pressure terms).  Note that the $\Gamma_e$ values in
Eq.~(\ref{eq:gibbsion}) are evaluated at $V=V_0$, while those in
Section~\ref{sub:tcp} are evaluated at $\simeq V_{01}$. In order to
show the two sets of phase diagrams on the same axis we use the
relation [cf. Eq.~(\ref{eq:deltaV})]:
\bal
\Gamma_e(V_{01}) = {}& \Gamma_e(V_0)\left(\frac{V_{01}}{V_0}\right)^{1/3} \nonumber\\
= {}& \Gamma_e(V_0)\left[1 + \frac{2\alpha}{(18\pi)^{1/3}\langle Z \rangle}\frac{\sqrt{1+y^2}}{y} \frac{\partial f_l}{\partial \Gamma_e}\right]^{1/3} \,,
\label{eq:gammaV01}
\eal
where all instances of $y$ and $\Gamma_e$ on the right-hand side of
Eq.~(\ref{eq:gammaV01}) are evaluated at $V=V_0$. Here we choose to
solve for $\Gamma_e(V_{01})$ of the liquid, although the results are
practically the same if $\Gamma_e(V_{01})$ of the solid is used
instead (since the two $\Gamma_e$ values differ by at most $0.004\%$
even for $R_Z \simeq 4$). Our results, plotted as a function of
$\Gamma_1(V_{01})=Z_1^{5/3}\Gamma_e(V_{01})$, are shown in
Fig.~\ref{fig:gibbscomp}. Not surprisingly, we obtain results very
similar to those found by \cite{ogata93}: the assumption of
transitions at constant volume rather than at constant pressure has no
effect on the phase diagram unless $R_Z \agt 2$, in which case the
only effect is to widen the unstable region slightly. For $2 < R_Z <
5$ the unstable region widens by at most $1$-$2\%$, with the largest
change occurring for $\Gamma_1 \alt \Gamma_{\rm crit}$. Since the
calculation of Section~\ref{sec:results} was done at a relatively low
value of $\Gamma$ (at $\Gamma_{Z=8} \simeq 27$, which is below
$\Gamma_{\rm crit}$ for all species $Z < 25$), we expect that the
results shown there will not change when the Gibbs free energy is
used. At large $\Gamma$, however, when nearly all of the mixture is in
the solid state (see Section~\ref{sec:discuss}), inclusion of the
Gibbs free energy in the equations of Section~\ref{sub:mcp} may be
necessary to accurately determine the phase transition properties
under these conditions.

\section{The deviation from linear mixing in the liquid}
\label{sec:delfl}

In our calculation we assume perfect linear mixing in the liquid
state, by setting $\Delta f_l=0$. We discuss here how our results
(Section~\ref{sec:results}) change when a more accurate form for
$\Delta f_l$ is used.

There are several fitting formulae of $\Delta f_l$ available in the
literature (e.g, Refs.~\cite{ogata93,dewitt96,potekhin09b}). We choose
to implement the fit from Equation~(9) of \citet{potekhin09b}
(hereafter PCCDR), since it provides accurate results for $\Delta f_l$
over a wide range of $\Gamma$ values, $Z$ ratios, and fractional
abundances of each species. It is also the only fit we are aware of
that is immediately applicable to plasmas with more than two
components, though we do not make use of that feature here.

\begin{figure}
\begin{center}
\includegraphics[width=\columnwidth]{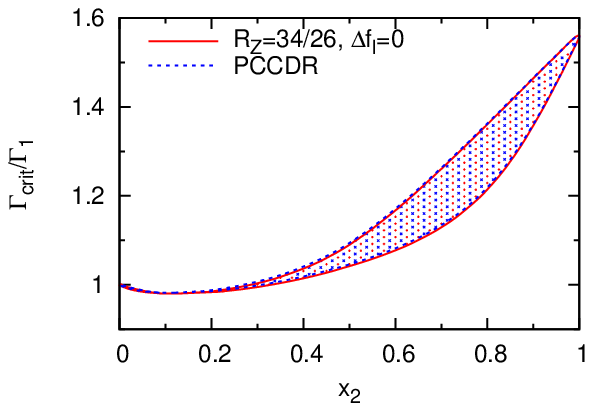}
\includegraphics[width=\columnwidth]{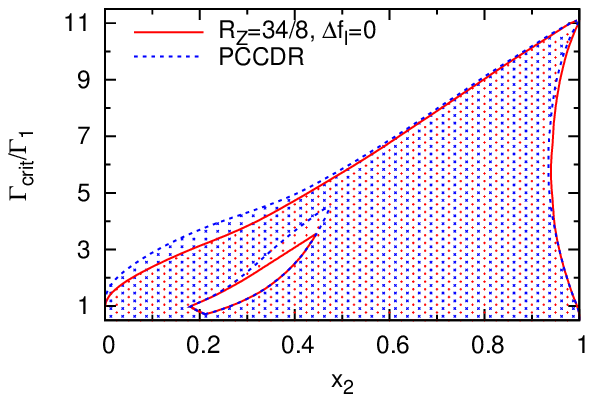}
\end{center}
\caption[Phase diagrams with and without a liquid deviation term]
{(Color online) Phase diagrams for charge ratios $R_Z = 34/26$ (top
panel) and $R_Z = 34/8$ (bottom panel). Phase transitions where the
liquid deviation term $\Delta f_l$ is ignored are labeled ``$\Delta
f_l=0$'', and transitions where the liquid deviation is given by
Eq.~(9) of \citet{potekhin09b} are labeled ``PCCDR''.}
\label{fig:delflcomp}
\end{figure}

We calculate the phase diagrams for two-component plasmas with charge
ratios $R_Z$ up to $34/8$, first for $\Delta f_l=0$, and then using
Eq.~(9) of PCCDR (i.e., for $\Delta f_l \ne 0$). Our results are shown
in Fig.~\ref{fig:delflcomp}. We find that the assumption $\Delta
f_l=0$ has no effect on the phase diagram unless $R_Z \agt 3$, in
which case the only effect is to shift the low-$x_2$ side (the left
side, in Fig.~\ref{fig:delflcomp}) of the unstable region toward even
smaller values of $x_2$. The shift is most significant for large $R_Z$
and $\Gamma$, with shifts of around $5\%$ of the width of the unstable
region for $R_Z \simeq 4$ and $\Gamma_1 \simeq \Gamma_{\rm
crit}$. Since our calculation was done at a relatively low value of
$\Gamma$, we expect that the results of Section~\ref{sec:results} will
not change when an accurate form for $\Delta f_l$ is used
(cf. Section~\ref{sec:gibbs}). At larger values of $\Gamma$, a $\Delta
f_l$ term may be necessary to ensure the accuracy of the calculation.

Here and in Section~\ref{sec:gibbs} we have compared phase diagrams
generated by our calculation to those that are generated if additional
terms are considered. We can also compare our phase diagrams to those
of other works. Particular fruitful comparisons can be made with
\citet{segretain93} and \citet{ogata93}, since these works present
phase diagrams at several different values of $R_Z$; the $R_Z$ values
in Figs.~\ref{fig:gibbscomp} and \ref{fig:delflcomp} were chosen in
part because of the similarity to the ratios presented in these two
works (i.e., $R_Z=34/26 \simeq 4/3 = 1/0.75$, $R_Z=34/20 \simeq 5/3
\simeq 1/0.55$, and $R_Z=34/8 \simeq 13/3$). Our diagrams agree
closely with those of \cite{ogata93}, with one important exception:
for most values of $R_Z$, this group finds `azeotropic points' or
eutectic points at $x_2 \alt 0.04$ that do not exist in our
diagrams. The close agreement for $x_2 > 0.04$ is due to the fact that
both our group and theirs used fitting formulae with the same form for
$\Delta f_s$ [Eq.~(\ref{eq:delusTCP}], while the poor agreement at
$x_2 < 0.04$ is due to the fact that we used $\Delta f_l=0$ while
\cite{ogata93} used a form for $\Delta f_l$ that was negative for $x_2
\alt 0.05$. Our diagrams agree less closely with those of
\cite{segretain93}, though the agreement is still very good at small
$\Gamma$ (in the upper half of each diagram). Even at large $\Gamma$
the diagrams of our group and theirs are qualitatively similar, with
the main differences being the larger amount of stable solid regions
at high $x_2$ and the delayed (in terms of increasing $R_Z$)
transition from spindle type to azeotropic type in the diagrams of
\cite{segretain93}. We find that the transition from spindle-type to
azeotropic-type phase diagrams occurs at $R_Z \simeq 1.2 \simeq 28/34
\simeq 1/0.83$, which is a somewhat lower value of $R_Z$ than found by
\citet{segretain93} or \citet{dewitt96} ($1/0.72 \simeq 1.4$).

\section{The deviation from linear mixing in the solid}
\label{sec:delfs}

In this section we provide a simple estimate of $\Delta f_s$ for
multi-component plasmas, using the approximation that only nearest
neighbors contribute to the interaction energy of each ion (see, e.g.,
Ref.~\cite{gordon68}). The expression found here is too simplistic for
use in our calculation, but illustrates the general form of $\Delta
f_s$ for plasmas with three or more components; the $\Delta f_s$ term
of Section~\ref{sec:method} [Eq.~(\ref{eq:delusMCP})] has a very
similar form.

Let $u_{ij} = U_{ij}/(Nk_BT)$ be the interaction energy between
nearest-neighbor ions of species $i$ and $j$ ($u_{ij} = u_{ji}$). When
all ion species are completely separated, the interaction energy per
ion for species $i$ is $u_{ii}/2$, and the total interaction energy of
the system is given by
\be
u_{\rm sep} = \frac{1}{2}\sum x_i u_{ii} \,.
\ee
When the ion species are mixed, the interaction energy per ion for
species $i$ is $\sum_j x_j u_{ij}/2$, assuming that the various ions
are randomly distributed throughout the mixture. The total energy of
the system is then
\be
u_{\rm mix} = \frac{1}{2}\sum_i \sum_j x_i x_j u_{ij} \,.
\ee
The internal energy of mixing for the solid, $\Delta u_s = u_{\rm
mix}-u_{\rm sep}$, is given by
\bal
\Delta u_s = {}& \frac{1}{2} \sum_i x_i \left(\sum_j x_j u_{ij} - u_{ii}\right) \nonumber\\
= {}& \frac{1}{2} \sum_i x_i \sum_{j \ne i} x_j \left(u_{ij} - u_{ii}\right) \nonumber\\
= {}& \sum_i \sum_{j > i} x_i x_j \left(u_{ij} - \frac{u_{ii}+u_{jj}}{2}\right) \,.
\eal
The free energy of mixing can be found from the thermodynamic identity
\be
f = \int_0^\beta \frac{u(\beta')}{\beta'} d\beta' \,,
\ee
where $\beta = 1/(k_BT)$ (see, e.g., Ref.~\cite{potekhin00}). Assuming
that the interaction energies $u_{ij}$ scale linearly with $\beta$
(which is true, e.g., if $u_{ij} \propto \Gamma_e$), we have
\be
\Delta f_s = \sum_i \sum_{j > i} x_i x_j \left(u_{ij} - \frac{u_{ii}+u_{jj}}{2}\right) \,,
\ee
which of the same form as Eq.~(\ref{eq:delusMCP}).

\end{document}